\documentstyle[12pt]{article}
\newcommand{\bce}{\begin{center}}
\newcommand{\ece}{\end{center}}
\newcommand{\beq}{\begin{equation}}
\newcommand{\eeq}{\end{equation}}
\newcommand{\bea}{\vspace{0.25cm}\begin{eqnarray}}
\newcommand{\eea}{\end{eqnarray}}

\newcommand{\ba}{\begin{array}}
\newcommand{\ea}{\end{array}}

\newcommand{\doublespace}{
    \renewcommand{\baselinestretch}{1.6}\large\normalsize}

\def\lsim{\mathrel{\rlap{\lower4pt\hbox{\hskip1pt$\sim$}}
    \raise1pt\hbox{$<$}}}	  
\def\gsim{\mathrel{\rlap{\lower4pt\hbox{\hskip1pt$\sim$}}
    \raise1pt\hbox{$>$}}}	  

\def\Pom{{\bf I\!P}}

\def\lsim{\mathrel{\rlap{\lower4pt\hbox{\hskip1pt$\sim$}}
    \raise1pt\hbox{$<$}}}         
\def\gsim{\mathrel{\rlap{\lower4pt\hbox{\hskip1pt$\sim$}}
    \raise1pt\hbox{$>$}}}         

\def\Pom{{\bf I\!P}}

  \advance\oddsidemargin  by -1in
  \advance\evensidemargin by -1in
  \advance\topmargin      by -0.5in
\def\lsim{\mathrel{\rlap{\lower4pt\hbox{\hskip1pt$\sim$}}
    \raise1pt\hbox{$<$}}}         
\def\gsim{\mathrel{\rlap{\lower4pt\hbox{\hskip1pt$\sim$}}
    \raise1pt\hbox{$>$}}}         

\def\Pom{{\bf I\!P}}
\def\beq{\begin{equation}}
\def\endeq{\end{equation}}
\def\arr{\begin{eqnarray}}
\def\endarr{\end{eqnarray}}
\makeindex

\doublespace

\begin{document}


\phantom{.}{\bf \Large \hspace{10.0cm} KFA-IKP(Th)-1994-6 \\
\phantom{.}\hspace{11.9cm}24 January   1994\vspace{0.4cm}\\ }

\begin{center}
{\bf\sl \huge Deep inelastic scattering at HERA
and the BFKL pomeron}
\vspace{0.4cm}\\
{\bf \large
N.N.~Nikolaev$^{a,b}$, and B.G.~Zakharov$^{b}$
\bigskip\\}
{\it
$^{a}$IKP(Theorie), KFA J{\"u}lich, 5170 J{\"u}lich, Germany
\medskip\\
$^{b}$L. D. Landau Institute for Theoretical Physics, GSP-1,
117940, \\
ul. Kosygina 2, Moscow 117334, Russia.
\vspace{1.0cm}\\ }
{\Large
Abstract}\\
\end{center}
We present
a BFKL phenomenology of structure functions in the
framework of the generalized BFKL equation for the dipole
cross section proposed by us recently. We show that the
HERA measurements of
excitation of charm at $Q^{2} \lsim $10 GeV$^{2}$ allow
determination of the intercept of the BFKL pomeron.
A comparison with the
experimental data suggests a large contribution from the
nonperturbative cross section for colour dipoles of large
size $r$.
 \bigskip\\

\begin{center}
E-mail: kph154@zam001.zam.kfa-juelich.de
\end{center}

\pagebreak

The $S$-matrix of diffractive scattering is exactly diagonalized
in terms of the dipole cross section [1-3] which satisfies the
generalized BFKL (Balitskii-Fadin-Kuraev-Lipatov)
equation derived in [4-7]. In [6,7] we have determined the
intercept of the pomeron $\Delta_{\Pom}$
and found the pomeron dipole cross
section
in the realistic model with the running
QCD coupling $\alpha_{S}(r)$
and the finite correlation radius $R_{c}$ for
perturbative gluons. With the $\mu_{G}=1/R_{c}=0.75$GeV and the
infrared freezing of the strong coupling $\alpha_{S}(r \geq R_{f})=
\alpha_{S}^{(fr)}=0.8$ we found $\Delta_{\Pom}=0.4$ [6].

In this paper we discuss the coresponding
BFKL phenomenology of deep inelastic
scattering at large ${1\over x}$, where $x$ is the Bjorken variable.
The starting point of our analysis is the dipole-cross section
representation for the virtual photoabsorption cross section [1]
\beq
\sigma_{T,L}(\gamma^{*}N,\xi,Q^{2})=
\int_{0}^{1} dz\int d^{2}\vec{r}\,\,
\vert\Psi_{T,L}(z,r)\vert^{2}\sigma(\xi,r)\, ,
\label{eq:1}
\endeq
where $\sigma(\xi,r)$ is the dipole cross section for the
colour dipole of size $r$, $Q^{2}$ is the virtuality of the photon,
$\xi=\log({x_{0}\over x})$ and
$x_{0}\sim $0.1-0.01 corresponds
to the onset of the leading-$\log({1\over x})$ approximation.
In the perturbative part of the dipole cross
section $\sigma^{(pt)}(\xi,r)$, which
is a solution of our generalized BFKL equation, the gluon
correlation radius $R_{c}$ emerges as the principal parameter
which controls its $r$ dependence and the absolute normalization.
At $r> R_{c}$ the dipole cross section contains also the
nonperturbative component, which is of major concern from the
point of view of the phenomenology to be presented in this paper.
The wave functions of the (T) transverse
and (L) longitudinal virtual
photon were derived in [1] and read
\beq
\vert\Psi_{T}(z,r)\vert^{2}={6\alpha_{em} \over (2\pi)^{2}}
\sum_{1}^{N_{f}}e_{f}^{2}
\{[z^{2}+(1-z)^{2}]\varepsilon^{2}K_{1}(\varepsilon r)^{2}+
m_{f}^{2}K_{0}(\varepsilon r)^{2}\}\,\,,
\label{eq:2}
\endeq
\beq
\vert\Psi_{L}(z,r)\vert^{2}={6\alpha_{em} \over (2\pi)^{2}}
\sum_{1}^{N_{f}}4e_{f}^{2}\,\,
Q^{2}\,z^{2}(1-z)^{2}K_{0}(\varepsilon r)^{2}\,\,,
\label{eq:3}
\endeq
where $K_{1}(x)$ is the modified Bessel function,
$e_{f}$ is the quark charge in units of the electron charge,
$\varepsilon^{2}=z(1-z)Q^{2}+m_{f}^{2}$,
$m_{f}$ is the quark mass and $z$ is
the fraction of photon's (lightcone)
momentum carried by the quark of the $q\bar{q}$ pair
($0 <z<1$). Then, the sea structure function is calculated as
$ F_{2}(x,Q^{2})=Q^{2}
\left[\sigma_{T}+\sigma_{L}\right]/(4\pi^{2}\alpha_{em})$.
The dipole cross section $\sigma(\xi,r)$ is universal, the
$Q^{2}$ and the flavour dependence of structure functions
only comes from $|\Psi_{T,L}|^{2}$.

The derivation [1] of (\ref{eq:1}) uses the observation that at
${1\over x} \gg 1$, the virtual photoabsorption
can be viewed as interaction with the target proton of
the multipartonic lightcone Fock states
($q\bar{q},\, q\bar{q}g...$) of the
photon, which are formed at large distance $\Delta z \sim
{1\over m_{p}x}$ upstream the target.
In [4] we gave the regular procedure for construction
of the corresponding multiparton lightcone wave function
and of the multiparton cross section.
The Fock states with $n$ soft gluons give the
$\propto \xi^{n}$ contribution to the total
photoabsorption cross section, which can be reabsorbed into the
energy dependent perturbative dipole cross section
\beq
\sigma^{(pt)}(\xi,r)=\sum_{n=0}{1\over n!}\sigma_{n}(r)
\xi^{n}\, ,
\label{eq:4}
\endeq
which satisfies the
generalized BFKL equation [2-5]
\beq
{\partial \sigma^{(pt)}(\xi,r) \over \partial \xi} ={\cal K}\otimes
\sigma^{(pt)}(\xi,r)\, ,
\label{eq:5}
\endeq
where in terms of the expansion (\ref{eq:4})
the kernel ${\cal K}$ is defined by
\arr
\sigma_{n+1}(r)={\cal K}\otimes\sigma_{n}(r)=~~~~~~~~~~~~~~~~~~~~~~~~~~~~
\nonumber\\
{3 \over 8\pi^{3}} \int d^{2}\vec{\rho}_{1}\,\,
\mu_{G}^{2}
\left|g_{S}(R_{1})
K_{1}(\mu_{G}\rho_{1}){\vec{\rho}_{1}\over \rho_{1}}
-g_{S}(R_{2})
K_{1}(\mu_{G}\rho_{2}){\vec{\rho}_{2} \over \rho_{2}}\right|^{2}
[\sigma_{n}(\rho_{1})+
\sigma_{n}(\rho_{2})-\sigma_{n}(r)]   \, \, .
\label{eq:6}
\endarr
Here $R_{c}=1/\mu_{G}$ is the correlation radius for perturbative
gluons, $\vec{\rho}_{2}=\vec{\rho}_{1}-\vec{r}$,
 $R_{i}={\rm min}\{r,\rho_{i}\}$,
$g_{S}(r)$ is the effective colour charge,
\beq
\alpha_{S}(r)={g_{S}(r)^{2}\over 4\pi}={4\pi \over
\beta_{0}\log\left(
{C^{2} \over \Lambda_{QCD}^{2}r^{2}}\right)} \, ,
\label{eq:7}
\endeq
where $\beta_{0}=11-{2\over 3}N_{f}=9$ for $N_{f}=3$
active flavours, $C\approx 1.5$ [1] and we impose the infrared
freezing $\alpha_{S}(r>R_{f})=\alpha_{S}^{(fr)}=0.8$ [6].
The correspondance to the original BFKL equation [8] was
discussed in detail in [4-7].

Once the dipole cross section (\ref{eq:4})
is known, Eq.~(\ref{eq:1}) enables
the parameter-free calculation of the proton structure function at
small $x$, which is the subject of the present communication.
Below
we discuss the salient features of the emerging BFKL phenomenology
of deep inelastic scattering at small $x$.
We start with the discussion of the nonperturbative component of
the dipole cross section, which dominates at large $r$ and at
moderately large ${1\over x}$.
Then, we present calculations of the proton
structure function in the kinematical range of HERA and find good
agreement with the data from ZEUS and H1 experiments [9,10]. We
comment on determination of the pomeron intercept from the
charm structure function. We also
give an estimate of the gluon distributions which correspond to
these structure functions and give predictions for
$R=\sigma_{L}/\sigma_{T}$.

We start presentation of our results with the description of the
colour dipole cross secton. In the perturbative QCD, the growth
of the dipole cross section comes from the logarithmic growth of
the multiplicity $n_{g}$ of perturbative guons in the photon
generated from the parent $q\bar{q}$ Fock state
$
n_{g} \propto \int_{x}^{x_{0}} {dz_{g}/ z_{g}} =\xi.$
Equations (\ref{eq:5},\ref{eq:6})
describe the perturbative part of the dipole cross section:
the kernel ${\cal K}$ is
directly proportional to the probability of radiation of the
perturbative gluons [4-7], and
$
\Delta\sigma_{g}={9\over 8}[\sigma(\rho_{1})+
\sigma(\rho_{2})-\sigma(r)]
$
shows how much the interaction cross section is changed for the
presense of the perturbative gluon (here
$\vec{\rho}_{1,2}$ is the separation of the gluon from the
(anti)quark,
$\vec{\rho}_{2}=
\vec{\rho}_{1}-\vec{r}$ and $\sigma(r)$ is the dipole cross section
for the parent $q\bar{q}$-dipole of size $\vec{r}$).
To higher orders, the gluons themselves become the source
of the subsequent generation of gluons, so that Fock states with
$n$ soft gluons give $\propto \xi^{n}$ contribution to the dipole
cross section.
As a boundary conditon at $x_{0}=3\cdot 10^{-2}$,
one can start with the
two-gluon exchange
dipole cross section for interaction with the nucleon target [1]
\beq
\sigma_{0}(r)=
{32 \over 9}
\int {d^{2}\vec{k}
\over(\vec{k}^{2}+\mu_{G}^{2})^{2} }
\alpha_{S}(k^{2})\alpha_{S}(\kappa^{2})
[1-G_{p}(3\vec{k}^{2})]
\left[1-\exp(i\vec{k}\vec{r})\right] \, ,
\label{eq:8}
\endeq
where $G_{p}(q^{2})$ is the charge form factor of the proton,
$\kappa^{2}={\rm max}\{k^{2},{C^{2}\over r^{2}}\}$.

The rightmost singularity in the complex $j$-plane corresponds to
the pomeron cross section
$
\sigma_{\Pom}(\xi,r)=\sigma_{\Pom}(r)\exp(\Delta_{\Pom}\xi) \, .
$
The crucial point of the BFKL phenomenology is that the pomeron
cross section $\sigma_{\Pom}(r)$ is not, and can not, be confined
to the small-$r$ region, which is due to the so-called diffusion
property of the Green's function of the BFKL equation [8]. The
underlying physics is very simple: even when one starts with
the small-size beam and target colour dipoles, radiated gluons
stick out of the parent dipoles the transverse distance $\sim R_{c}$
apart, and it is interaction of these gluons which controlls the
total cross section. For instance, the pomeron
intercept $\Delta_{\Pom}$ is predominatly controlled
by the behavior of $\sigma_{\Pom}(r)$ in the semi-perturbative
region of $r\sim R_{c}$ [6]. Consequently,
the contribution from the infrared region of large $r$, and the
infrared regularization of the perturbative QCD, are at the heart
of the BFKL phenomenology.

Motivated by the lattice studies of the correlation function of
perturbative gluons [11], we take $\mu_{G}=0.75$GeV. With such
a large value of $\mu_{G}$, the driving term (\ref{eq:8}) of the
perturbative dipole cross section $\sigma^{(pt)}(\xi=0,r)
=\sigma_{0}(r)$ flattens
at large $r$ at $\sim 8$mb (Fig.~1). One can easily anticipate
the nonperturbative mechanism of interaction of large,
$r \gsim R_{c}$, colour dipoles.
Here we make an assumption that the perturbative
and nonperturbative cross sections are additive [12]
\beq
\sigma(\xi,r)=\sigma^{(pt)}(\xi,r)+\sigma^{(npt)}(r)\, .
\label{eq:9}
\endeq
Because the rise of the $\sigma^{(pt)}(\xi,r)$
is due to the rising multiplicity of the perturbative
gluons in the photon wave function, we make a plausible assumption
that the nonperturbative dipole cross section does not rise
with energy. We also assume the dominance of the perturbative cross
section, $\sigma^{(npt)}/\sigma^{(pt)}(r) \ll 1$ at $r\ll R_{c}$.
In Fig.~1 we show our choice of $\sigma^{(npt)}(r)$.

Different processes probe $\sigma(\xi,r)$ at different $r$, which
allows to disentangle the perturbative and nonperturbative dipole
cross sections experimentally. For instance, we shall see that
the charm production
predominantly probes $\sigma^{(pt)}(\xi,r)$.
The above choice of $\sigma^{(npt)}(r)$ is already strongly
constrained by various experimental data. As a matter of fact,
we require that the sum of the perturbative and nonperturbative
cross sections $\sigma(\xi=0,r)$
roughly reproduces the colour dipole cross section
of Ref.~1, which was constrained to reproduce  $\sigma_{tot}(\pi N)
\approx 25$mb. The cross section of Ref.~1 has already been
succesfully used in the calculation of the structure functions
at moderate $Q^{2}$ (see also below)
and of the nuclear shadowing [1,3], of the
total photoproduction cross section and of the rate of the
diffraction dissociation of real and virtual photons [2].
The E665 data on the exclusive leptoproduction of the
$\rho^{0}$-mesons on nuclei [13] and the NMC data on the
$J/\Psi$ photoproduction on nuclei [14]
imply that $\sigma(\xi=0,r)$
decreases from $\sim 30$mb at $r\sim $(1-1.5)f to $\sim 6$mb
at $r\sim 0.4$f (for the detailed discussion see [15,16]).

Because the ratio $\sigma(\xi,r)/r^{2}$ only slowly increases towards
small $r$, it is
useful to rewrite Eq.~(\ref{eq:1}) in the form
\beq
F_{2}(x,Q^{2}= {1\over \pi^{3}}
\int {d r^{2} \over r^{2}}\cdot
{\sigma(\xi,r)\over r^{2}}\Phi_{T}(Q^{2},r^{2}) \, .
\label{eq:10}
\endeq
The divided by $e_{f}^{2}$
contributions of light and charmed quarks to the kernel
$\Phi_{T}(Q^{2},r^{2})=$ ($\pi^{2}/4\alpha_{em} )
\int_{0}^{1} dz \, Q^{2}r^{4}
\vert\Psi_{T,L}(z,r)\vert^{2}$ are shown in Fig.~2. They develope
a plateau at large $Q^{2}$, the emergence of which signals the
onset of the GLDAP (Gribov-Lipatov-Dokshitzer-Altarelli-Parisi) [17]
evolution of parton densities.
The Fig.~2 in conjunction with Fig.~1 shows which region of $r$
is probed in deep inelastic scattering at the virtuality $Q^{2}$.
Following [1-3], we take
the quark masses $m_{u,d}=150$MeV,\,$m_{s}=300$MeV,\, $m_{c}=
1.5$GeV,\,$m_{b}=4.5$GeV.
The structure functions are insensitive to $\sigma(\xi,r)$ at
$r \gsim 2$f. Notice, that the contribution from
$r\sim 1$f is the perfectly scaling part
of $F_{2}(x,Q^{2})$ which persists at all the values of $Q^{2}$ [1].
Numerically, we find that the contribution from $\sigma^{(npt)}(r)$
to $F_{2}(x,Q^{2})$ changes form
$F_{2}^{(npt)}(x,Q^{2}=0.75{\rm GeV}^{2})\approx 0.1$ to
$F_{2}^{(npt)}(x,Q^{2}=15{\rm GeV}^{2})\approx 0.13$.
In Fig.~3a we show the total sea structure function, in Fig.~3b
we show its perturbative part.
Because of
expansion of the plateau in Fig.~2 with rising $Q^{2}$, the
contribution to $F_{2}(x,Q^{2})$ from $\sigma^{(pt)}(\xi,r)$
takes over at large $Q^{2}$.
The values of structure function already at $Q^{2} \gsim
$(2-4)\,GeV$^{2}$ are insensitive to masses of light quarks.
Contribution from $\sigma^{(pt)}(\xi,r)$ also takes over
at large ${1\over x}$, but at $x\sim 10^{-2}$ the effect of
$F^{(npt)}(x,Q^{2})$ is quite substantial.

In Fig.~4 we compare our predictions with the NMC [18], ZEUS [10]
and H1 [11] structure functions. The overall agreement is good;
the slight
departure from the NMC data points at large $x$ is due to the
the valence structure function not considered here. Besides,
at large $x$ the nonperturbative cross section can contain the
Regge-decreasing component
$\propto \exp(-{1\over 2}\xi)=\sqrt{x/x_{0}}$, for the magnitude
of which we do not have any educated guess [19].
It is somewhat more accurate to use in
Eq.~(\ref{eq:1}) $\sigma(\xi+\log z,r)$, but for the purposes
of the qualitative phenomenology of the present paper,
this rescaling of
the energy variable can be reabsorbed into renormalization of
the boundary condition.

In [7] we have shown that
the energy dependence of $\sigma^{(pt)}(\xi,r)$ at $r\sim 0.5 R_{c}$
follows the asymptotic law $\propto \exp(\Delta_{\Pom}\xi)$
starting already at moderate energy. At $r \ll R_{c}$ the
effective intercept is substantially larger than $\Delta_{\Pom}$
beacuse of the double-leading-logarithm effects [5,7].
Eq.~(15) and Fig.~2 show that the total structure function
$F_{2}(x,Q^{2})$ receives contributions from a broad range of $r$.
Besides, Figs.~3a and 3b show that the $x$-dependence
of $F_{2}(x,Q^{2})$ is strongly
affected by the contribution from the
nonperturbative cross section, which
does not allow the model-independent
determination of the pomeron intercept $\Delta_{\Pom}$
form the total structure fucntion measurements
in the kinematical range of HERA.
However, the same Fig.~2 shows that one can zoom
at the perturbative cross section at
$r\sim $0.15f measuring excitation of charm at $Q^{2} \lsim
10^{2}$, so that
in this case the effective intercept
$\Delta_{eff}(x,Q^{2})=
-{\partial \log F_{2}(x,Q^{2})/ \partial \log x}
\approx  \Delta_{\Pom}= 0.4$ . At larger $Q^{2}$,
$\Delta_{eff}(x,Q^{2})$ overestimates $\Delta_{\Pom}$ because
of the double-leading-logarithm efects [7].
Our predictions for the charm
structure function are shown in Fig.~3c (The threshold effects
in excitation of heavy flavours at small $Q^{2}$ [20] can
approximately be taken into account by the rescaling
$\xi \Longrightarrow \xi + \log(1+{m_{c}^{2}\over Q^{2}})$).
Once the pomeron intercept and the perturbative dipole cross
section are determined, one can separate the nonperturbative
contribution from the high-accuracy data on
$F_{2}(x,Q^{2})$ at relatively low $Q^{2}\lsim $20\,GeV$^{2}$.
To this end, the important point is that at $Q^{2}\gsim$
(10-20)\,GeV$^{2}$ the BFKL evolution and the conventional
GLDAP (Gribov-Lipatov-Dokshitzer-Altarelli-Parisi) [17]
evolution complemented by the proper boundary condition,
yield essentially identical predictions for the
$F_{2}^{(pt)}(x,Q^{2})$ in the HERA region (the deep reasons
for this similarity are discussed in detail in [7]).

Above we only presented the results for the bare pomeron. At large
${1\over x}$ the unitarization effects become important. They
were discussed to great detail in [4,21]. The unitarity bound
can best be formulated in terms of the profile function (the
partial wave amplitude) $\Gamma(\xi,r,b)$ for the dipole cross section
\beq
\Gamma(\xi,r,b)\approx {\sigma(\xi,r) \over 4\pi B_{0}(\xi,r)}
\exp\left[-{b^{2}\over 2B_{0}(\xi,r)}\right]\, ,
\label{eq:11}
\endeq
where $\vec{b}$ is the impact parameter. The diffraction slope
$B_{0}(\xi,r)$ is a slow function of $\xi$ and $r$
compared to the rapidly changing $\sigma(\xi,r)$.
Evidently, the $s$-channel unitarity constraint $\Gamma(\xi,r,b)<1$
will first be violated at large $r$.
To a good approximation,
the unitarization effects reduce to a (very weakly $Q^{2}$
dependent) renormalization factor $(1-\delta\xi)\sim
\exp(-\delta\xi)$
with $\delta \sim$0.03-0.05 [4,21], which is negligibly
small compared to $\Delta_{\Pom}=0.4$.
Furthermore, in [4] we have shown that the
unitarized structure function
still satisfies the linear GLDAP evolution. Consequently, for the
purposes of the present crude phenomenology, the unitarization
effects can be reabsorbed into the (as yet unknown)
intercept of the pomeron.

The ratio $R=\sigma_{L}/\sigma_{T}$ was first evaluated in the
framework of the dipole cross section technique in [1].
In [21] we have shown that the unitarization effects have
little impact on $\sigma_{L}/\sigma_{T}$.
Our predictions for $R$ for the sea component of the structure
function
are shown in Fig.~6. We find a sort of
crossover at $x\sim (1-2)\cdot 10^{-3}$. At small $x$, $R$ rises
a little with $Q^{2}$, and is essentially flat at $Q^{2}\gsim 20
$GeV$^{2}$.
The knoweledge of $R$ at small $Q^{2}$ is important for certain
applications like an accurate estimate
of the radiative corrections.
In Fig.~6b  we separately
show our predictions for $R$ at $Q^{2} \leq 0.75$
GeV$^{2}$. Notice that $|\Psi_{L}|^{2} \propto Q^{2}$, and
at very small $Q^{2}$ we have $R\propto Q^{2}$ just by
virtue of gauge invariance.

The relationship between our dipole-cross section representation
for deep inelastic scattering and the more
familar parton model (recall that the latter comes along with the
Weizs\"acker-Williams reinterpretation of the photoabsorption
cross section [17]) is established by [3,4,21,22]
\arr
\sigma(\xi,r)\approx
{4\pi \over 3}\alpha_{S}(r)\int {d^{2}\vec{k} \over k^{2}}
\cdot{1-\exp(i\vec{k}\vec{r}) \over k^{2}}\cdot
{dG(\xi,k^{2})\over d\log k^{2}}
 \approx
{\pi^{2}\over 3}r^{2}\alpha_{S}(r)
G_{pt}(\xi,Q^{2}\sim {1\over r^{2}})\, ,
\label{eq:12}
\endarr
where $G(\xi,Q^{2})=xg(x,Q^{2})$ is the gluon structure function at
$x=x_{0}\exp(-\xi)$. Then, assuming $N_{f}=4$ active flavours, the
differentiation of Eq.~(\ref{eq:10}) gives (see also [23])
\beq
{ \partial F_{2}(x,Q^{2}) \over \partial \log Q^{2}}=
\left(\sum_{u,d,s,c}e_{f}^{2}\right)
 {\alpha_{S}(Q^{2})\over 3\pi}
G(2x,Q^{2})\, ,
\label{eq:13}
\endeq
where $e_{i}$ is the quark charge in units of the electron charge.
Strictly speaking, Eq.~(\ref{eq:12}) was derived for the
perturbative dipole cross section,
but one may use Eq.~(\ref{eq:12})
in conjunction with Eq.~(\ref{eq:13}) as a definition of the
small-$x$ gluon stucture function beyond the perturbation theory.
If we apply the procedure (\ref{eq:13}) to our BFKL structure
functions, then we find $G(x,Q^{2})$ shown in Fig.~7 in comparison
with the recent H1 [24] and NMC [18] determinations of
$G(x,Q^{2})$.
We have a good overall agreement with the experiment.
\medskip\\
{\Large \bf Discussion of the results and conclusions:}
\smallskip\\
We presented the BFKL phenomenology of deep inelastic
scattering at large ${1\over x}$
in the dipole-cross section representation.
The nover features of our approach, which make it different
from the related work by other authors [25], are as follows:
Firstly, its irrefutable advantage is an exact
diagonalization of the $S$-matrix in the dipole-cross section
representation.
Secondly, the general idea, formulated and exploited in
our early works [1-3,20-22], was to use the universality
property of the dipole cross section to fix the
normalization and the shape of $\sigma(\xi,r)$ at moderately
large energies. Thirdly, the
the nonperturbative dipole cross section which emerges as an
integral part of the formalism, is
well constrained by the hadronic cross sections [1], the
small-$Q^{2}$ structure functions [1,3], the diffraction dissociation
of photons [2], the nuclear shadowing [3] and the leptoproduction
of vector mesons [15,16].
Fourthly, the discussion in [25] centered on matching
the GLDAP solutions with the fixed-$\alpha_{S}$ solutions of the
scaling BFKL equation [8]. Solutions of our generalized BFKL
equation, which consistently treats the running coupling
$\alpha_{S}(r)$,
are dramatically different from the fixed-$\alpha_{S}$ solutions.
(for the more detailed discussion see [7]). Finally, our infrared
regularization is consistent with gauge invariance [4-6]

In our earlier work  [1-3,20-22]
we tried to describe the whole dipole cross section
$\sigma(\xi,r)$ purely perturbatively. As far as the absolute
value of $\sigma(\xi=0,r)$ is concerned, that
can be achieved at the expense of small $\mu_{G}$. Such a
perturbative glue always starts with $g(x,Q^{2}) \propto
{1 \over x}$ and always generates the sea $\bar{q}(x,Q^{2})
\propto {1\over x}$, providing a nice continuity from the real
photoproduction to deep inelastic scattering. The trouble with
small $\mu_{G}$ is a too rapid a buildup of the perturbative
glue, which can be tamed, and a
good quantitative description of the ZEUS and H1 structure
functions can be obtained,
at the expense of starting the GLDAP evolution of glue at
$Q_{0}^{2}\approx 1$GeV$^{2}\gg 4\mu_{G}^{2}$ [21] (the unitraization
effects were also found important to bring the predictons [21]
closer to the ZEUS and H1 data).
There is much correspondance between
this large value of $Q_{0}^{2}$ of Ref.~21 and the small gluon
correlation radius $R_{c}$ used in the present paper.
The present analysis shows that in order to understand the
HERA data, one really needs a large nonperturbative dipole
cross besides the
perturbative BFKL cross section.

We conclude listing the uncertanties and challenges:
The pomeron intercept and the pomeron cross section are
evidently sensitive to the specific choice of the
gluon correlation radius $R_{c}$ and of the freezing coupling,
and also can change if more sophisticated infrared regularization
is introduced. Another uncertainty is the choice of $x_{0}$.
Hovever, none of the above will drastically modify the
salient features of the above presented
dipole-cross section phenomenology .
The further progress in the theory requires the
experimental determination of the pomeron intercept $\Delta_{\Pom}$,
which the HERA experiments on excitation of charm can do,
and better understanding of the nonperturbative cross section,
which also can be probed in the HERA experiments.
{}From the theoretical side, more work on the unitarization effects
is in order.
\bigskip\\
{\bf Acknowledgements}: B.G.Z. is grateful to
J.Speth for the hospitality at IKP, KFA J\"ulich, where this
work was initiated.
\pagebreak

{\bf Figure captions:}

\begin{itemize}
\item[Fig.1 - ]
The nonperturbative and perturbative components of the dipole
cross section.

\item[Fig.2 - ]
The kernel $\Phi_{T}(Q^{2},r^{2})$ for (the top box) the
light ($u,d$) and (the bottom box) the charmed quarks.

\item[Fig.3 - ]
The predicted BFKL structure functions at small $x$: (a) the total
structure function, (b) the perturbative part of the structure
function, (c) the charm structure function. The curves correspond
to (from bottom to top) $Q^{2}$=0.75,\,1.25,\,1.75,\,
2.5,\,3.5,\,4.5,\,8.5,\,\-15,\,\-
30,\,\-60,\,\-120,\,\-240,\,480 GeV$^{2}$,
respectively.

\item[Fig.4 - ]
Comparison of the predicted BFKL structure functions with
the data from the ZEUS [10], H1 [11] and NMC [18] experiments.

\item[Fig.5 - ]
Predictions for $R=\sigma_{L}/\sigma_{T}$. The lower box shows
the low-$Q^{2}$ behavior of $R$.

\item[Fig.6 - ]
Comparison of the gluon structure function evaluated
from the predicted
BFKL structure functions using Eq.~(\ref{eq:13}) with the
H1 [24] and NMC [18] determinations.

\end{itemize}
\end{document}